# An Intelligent AI glasses System with Multi-Agent Architecture for Real-Time Voice Processing and Task Execution

Sheng-Kai Chen[1], Jyh-Horng Wu[2], Ching-Yao Lin[3] and Yen-Ting Lin[4]

Division of Virtual-Real Integration and Digital Twins

National Center for High-Performance Computing

Hsinchu, Taiwan

s1134807@mail.yzu.edu.tw[1], {jhwu[2], chingyao[3]}@niar.org.tw, 2303131@niar.org.tw[4]

**ABSTRACT**

**This paper presents an AI glasses system that integrates real-time voice processing, artificial intelligence(AI) agents, and cross-network streaming capabilities. The system employs dual-agent architecture where Agent 01 handles Automatic Speech Recognition (ASR) and Agent 02 manages AI processing through local Large Language Models (LLMs), Model Context Protocol (MCP) tools, and Retrieval-Augmented Generation (RAG). The system supports real-time RTSP streaming for voice and video data transmission, eye tracking data collection, and remote task execution through RabbitMQ messaging. Implementation demonstrates successful voice command processing with multilingual support and cross-platform task execution capabilities.**

*Keywords: AI glasses, speech recognition, artificial intelligence, RTSP streaming, eye tracking, distributed systems*

## 1. INTRODUCTION

Artificial Intelligence (AI) glasses represent a transformative solution for hands-free applications across industrial automation, healthcare, and consumer electronics. However, current AI glasses face significant limitations in real-time voice processing, intelligent task interpretation, and reliable cross-network communication. These constraints limit practical deployment in enterprise environments where seamless voice interaction and remote task execution are critical requirements.

The integration of artificial intelligence with AR hardware presents unique challenges. Processing voice commands requires a low-latency speech recognition while maintaining accuracy across diverse acoustic environments. Intelligent task execution requires sophisticated natural language understanding and access to external data sources and computational resources. Network communication must support real-time streaming while adapting to varying connectivity conditions and security requirements.

This research addresses these challenges through a novel multi-agent architecture that splits speech recognition from AI-driven processing and meets real-time performance constraints. The system combines hardware-level eye tracking data collection and software-based voice processing to create a comprehensive AI interaction platform. The architecture enables distributed task execution across heterogeneous network environments and preserves low-latency communication requirements needed for real-world applications.

## 2. RELATED WORK

Voice-controlled AR systems have been explored in various contexts, primarily focusing on single-component solutions. Aouam et al. [1] demonstrated voice-activated AR interfaces for automobile assembly applications, where voice commands are recognized and translated into textual commands for AR system interpretation, though achieving limited accuracy in noisy environments and relying on cloud-based processing that introduces latency constraints for real-time applications.

Multi-agent architectures in distributed systems have shown promise for complex task processing. Fortino et al. [2] have proposed agent-oriented cooperative smart object frameworks for IoT system design and implementation, though their approach did not address either real-time constraints or voice processing requirements specific to AR applications.

RTSP streaming for AR applications has been investigated primarily for video transmission. Hosseini and Georganas [3] implemented enhanced distributed streaming systems based on RTP/RTSP with resurgent ability, focusing on video quality rather than integrated voice and eye tracking data transmission. Eye tracking integration in AR systems has concentrated on gaze-based interaction mechanisms. Park et al. [4] developed wearable augmented reality systems using gaze interaction in controlled environments without addressing mobile usage scenarios or network resilience requirements.

The current literature lacks comprehensive solutions that integrate voice processing, intelligent task execution, and cross-network communication within a unified AR system. This study fills this gap through a multi-agent approach that addresses these requirements simultaneously.

## 3. METHODOLOGY

1) System Architecture Design

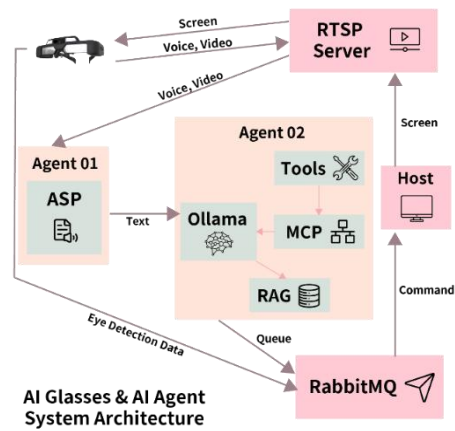

Figure 1 System Architecture

System architecture employs a distributed design comprising AI glasses hardware, processing agents, network communication infrastructure, and remote execution hosts. The design separates computational responsibilities across specialized components while maintaining coherent system operation through well-defined interfaces.

Figure 1 illustrates the overall system architecture. AI glasses capture voice, video, and eye tracking data transmitted via RTSP protocol to processing agents. Agent 01 handles speech recognition tasks while Agent 02 manages intelligent processing and task coordination. We employed RabbitMQ message queues to enable distributed task execution across remote hosts.

2) Agent 01: Speech Recognition Module

Agent 01 implements real-time speech recognition using Whisper.cpp framework [5] optimized for streaming audio processing. The module processes continuous audio streams through sliding window analysis with configurable buffer durations. Voice activity detection algorithms optimize processing resources by filtering silent audio segments.

Audio processing utilizes a sliding window approach where each buffer contains audio samples as defined in (1). The overlap ratio between consecutive windows is calculated according to (2) to ensure smooth processing transition. The overlap ensures smooth transitions between processing windows and prevents audio artifacts during segmentation. Voice Activity Detection applies energy and zero-crossing rate thresholds according to (3) to filter non-speech segments and optimize processing efficiency.

$$W_i = \{A(t) : t_{i-1} \leq t \leq t_i\} \quad (1)$$

where $A(t)$ is the continuous audio signal at time $t$,
and $t_i = t_0 + i \cdot \Delta t$ with $\Delta t$ being the window duration.

$$\alpha = \frac{t_{i-1} + \Delta t - t_i}{\Delta t} \quad (2)$$

where $\alpha$ represents the overlap ratio between consecutive windows,
$t_i$ is the start time of the current window,
$t_{i-1}$ is the start time of the previous window,
and $\Delta t$ is the window duration.

$$VAD(W_i) = \begin{cases} 1 & \text{if } E(W_i) > \theta_{energy} \text{ and } ZCR(W_i) < \theta_{zcr} \\ 0 & \text{otherwise} \end{cases} \quad (3)$$

where $E(W_i) = \sum_{t \in W_i} A(t)^2$ is the energy and $ZCR(W_i)$ is the zero-crossing rate.

The speech recognition pipeline begins with RTSP audio stream extraction using FFmpeg libraries [6] configured for minimal buffering. Audio preprocessing includes noise reduction filtering and automatic gain control. Multi-language support enables processing of English and Chinese speech inputs through language-specific acoustic models.

3) Agent 02: Intelligent Processing Framework

Agent 02 orchestrates complex reasoning tasks through integration of multiple AI subsystems. The framework combines local LLM processing using Ollama [7] with external tool access through Model Context Protocol (MCP) [8] and memory management via Retrieval-Augmented Generation (RAG).

Intent recognition utilizes a multi-stage analysis pipeline combining rule-based pattern matching with neural language model processing. Intent confidence scoring combines pattern matching and LLM outputs using weighted averaging as shown in (4). The pattern matching component uses substring matching metrics defined in (5), while contextual relevance incorporates conversation history through vector similarity measures.

$$C_{intent} = \omega_p \cdot P_s + \omega_l \cdot L_s + \omega_c \cdot C_{context} \quad (4)$$

where $\omega_p + \omega_l + \omega_c = 1$ and $C_{context}$ is the contextual relevance score.

$$P_s(c) = \max_{p \in \mathcal{P}} \frac{|match(c,p)|}{|p|} \quad (5)$$

where $match(c,p)$ returns the matching substring length for command $c$ against pattern set $\mathcal{P}$.

The RAG memory system implements ChromaDB [9] with sentence-transformers [10] for vector similarity search. Document retrieval utilizes cosine similarity computation between query and document embeddings as described in (6). The top-k retrieval function in (7) selects the most semantically relevant documents for context enhancement.

$$sim(\mathbf{q}, \mathbf{d}_i) = \frac{\mathbf{q} \cdot \mathbf{d}_i}{|\mathbf{q}||\mathbf{d}_i|} \quad (6)$$

where $\mathbf{q}$ is the query embedding vector,
$\mathbf{d}_i$ is the $i$-th document embedding vector,
$\mathbf{q} \cdot \mathbf{d}_i$ is the dot product,
and $|\mathbf{q}|, |\mathbf{d}_i|$ are the respective vector magnitudes.

$$R_k(\mathbf{q}) = \arg \max_{S \subset \mathbf{D}, |S|=k} \sum_{\mathbf{d}_i \in S} sim(\mathbf{q}, \mathbf{d}_i) \quad (7)$$

where $\mathcal{D}$ is the document corpus,
$S$ is a subset of documents with cardinality $k$,
and $sim(\mathbf{q}, \mathbf{d}_i)$ is the cosine similarity function.

4) Network Communication Protocol

The network architecture supports multiple connection methodologies to accommodate diverse deployment scenarios. Direct RTSP connections serve local area network environments, while port forwarding configurations enable wide area network access through NAT traversal mechanisms.

Network method selection employs multi-criteria optimization combining latency, bandwidth, and reliability metrics as formulated in Equation (8). The Network RTSP Manager implements automatic connection detection method through network topology analysis. Adaptive streaming rates adjust dynamically based on available bandwidth using the exponential decay function in Equation (9).

$$Score_{method} = \frac{\omega_1}{L + \epsilon} + \omega_2 \cdot B + \omega_3 \cdot R \quad (8)$$

where $L$ is latency, $B$ is bandwidth, $R$ is reliability, $\epsilon$ prevents division by zero, and $\omega_1 + \omega_2 + \omega_3 = 1$.

$$r(t) = r_{max} \cdot \min\left(1, \frac{B_{available}(t)}{B_{required}}\right) \cdot \exp(-\lambda \cdot L(t)) \quad (9)$$

where $\lambda$ is the latency penalty factor.

For enterprise deployments requiring secure communication, the system can be configured to operate over VPN connections. This capability addresses security requirements in corporate environments where voice and video data transmission must comply with organizational security policies. The network manager automatically detects VPN availability and adjusts streaming parameters to maintain performance within encrypted tunnels, ensuring that RTSP streaming and RabbitMQ messaging remain functional while meeting enterprise security standards.

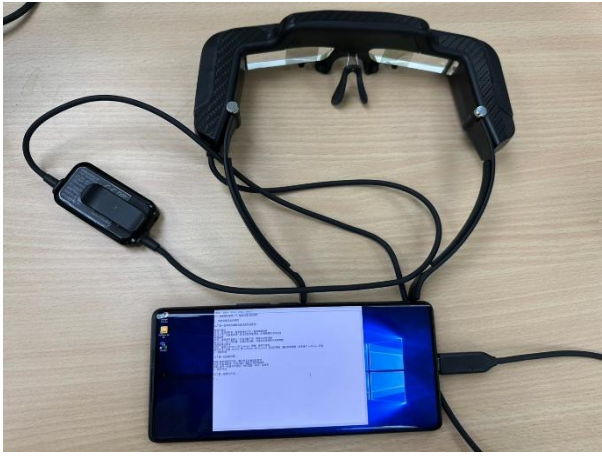

Figure 2 UR10 Results in AI Glasses

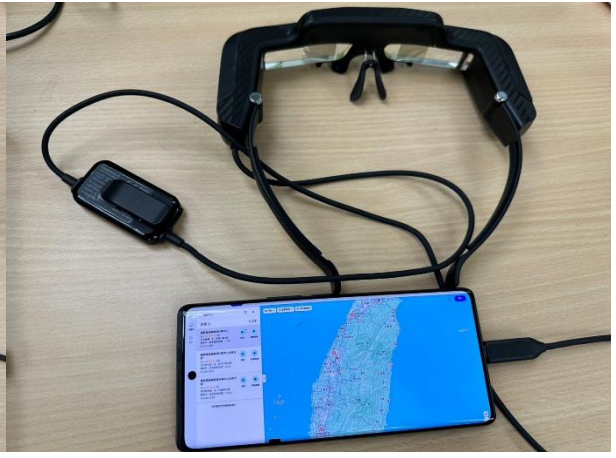

Figure 3 Map Results in AI Glasses

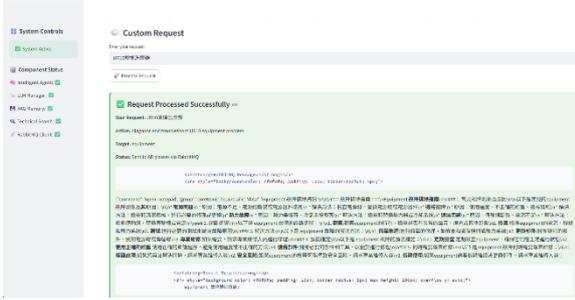

Figure 4 UR10 Request in Chatbot

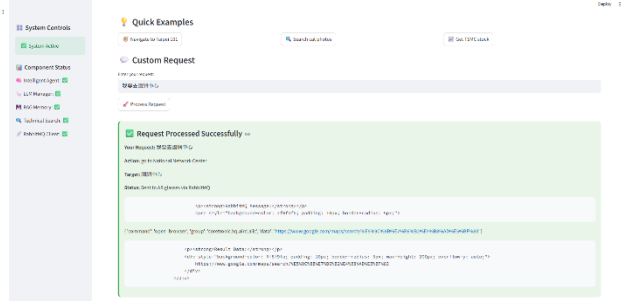

Figure 5 Map Request in Chatbot

Eye tracking data collection operates independently of voice processing to maintain temporal accuracy requirements. The Eye tracker Module interfaces with Ganzin hardware sensors through native Android APIs, capturing binocular gaze vectors at 30Hz sampling frequency.

Binocular gaze fusion applies confidence-weighted averaging as described in Equation (10) where individual eye weights depend on tracking confidence and noise characteristics according to Equation (11). World coordinate transformation uses calibrated rotation matrices and translation vectors as formulated in Equation (12).

$$\mathbf{g}_{combined} = \frac{w_L \mathbf{g}_L + w_R \mathbf{g}_R}{w_L + w_R} \qquad (10)$$

where $\mathbf{g}_L$ and $\mathbf{g}_R$ are left and right eye gaze vectors respectively, $w_L$ and $w_R$ are confidence-based weights, and $\mathbf{g}_{combined}$ is the fused gaze vector.

$$w_i = \frac{c_i^2}{c_i^2 + \sigma_i^2} \qquad (11)$$

where $c_i$ is the confidence score and $\sigma_i$ is the noise variance for eye $i$.

$$\mathbf{p}_{world} = \mathbf{R} \cdot \mathbf{g}_{eye} + \mathbf{t} \qquad (12)$$

where $\mathbf{R}$ is the calibrated per-user transformation matrix and $\mathbf{t}$ is the translation vector.

Real-time data transmission utilizes RabbitMQ messaging [11] with topic-based routing. Data serialization converts floating-point coordinates and timestamps to JSON format for network transmission with message persistence ensuring data integrity during network interruptions.

5) Distributed Task Execution

Task execution utilizes RabbitMQ message queuing to decouple command processing from task execution. The system implements topic-based routing to direct tasks to appropriate execution hosts based on task type and resource requirements.

Task scheduling employs priority-based queuing where task priority decreases exponentially with waiting time as defined in Equation (13). Algorithm 1 presents the task execution scheduling procedure that ensures resource constraints are respected while maximizing system throughput.

---
**Algorithm 1** Task Execution Scheduling
---
0: Initialize priority queue $Q$ **while** *system active* **do**

0:

$T \leftarrow$ highest priority task from $Q$ **if** *T.resources $\leq$ available resources* **then**

0:

Execute task $T$

0: Update resource allocation **else**

0:

Wait for resource availability

0:

0: =0
---

$$Priority(T_i) = \frac{U_i \cdot e^{-\alpha \cdot (t_{current} - t_{arrival})}}{D_i} \qquad (13)$$

where $U_i$ is the initial task utility,
$t_{current}$ and $t_{arrival}$ are current and task arrival times respectively,
$D_i$ is the decay factor for task $i$,
and the exponential term prevents task starvation.

The Task Executor framework provides cross-platform compatibility across Windows, macOS, and Linux environments. Task categories include web browser control, application launching, file system operations, and information display functions.

### 4. RESULTS

1) System Implementation

The system has been successfully implemented with all major components now operational. The dual-agent architecture effectively separates speech recognition from intelligent processing while preserving system coherence through well-defined messaging interfaces. Cross-network communication capabilities enable operation across diverse network infrastructures and automatically adapt to varying connectivity conditions.

2) Functional Validation

Figure 2 demonstrates the AI glasses display showing the result of UR10 robotic arm malfunction query. The system successfully renders comprehensive troubleshooting information on the AR interface, including equipment fault diagnosis and repair guidance in Chinese. The visual presentation includes structured diagnostic procedures and maintenance recommendations retrieved from the technical documentation system.

Figure 3 shows the AI glasses display presenting Google Maps location results for the navigation query. The system successfully renders the map interface showing the National Grid Center location with appropriate geographic markers and routing information. The AR display integrates seamlessly with the web-based mapping service to provide contextual location data.

Figure 4 presents the chatbot interface demonstrating successful RabbitMQ message transmission for the UR10 technical support request. The interface shows the system processing the equipment malfunction query and generating the appropriate RabbitMQ command message for transmission to AI glasses. The message structure includes command type, group routing, and comprehensive technical data payload.

Figure 5 displays the chatbot interface showing RabbitMQ message generation for the location request. The system demonstrates successful processing of the navigation command "I want to go to NCHC" and creation of the corresponding RabbitMQ message containing the Google Maps URL and routing information for AI glasses display.

3) Multi-Language Processing

The system demonstrates effective multilingual processing capabilities, which interpreting both English and Chinese voice commands accuaurately. Navigation requests, technical support queries, and general information requests are processed accurately in either language. The intent recognition system maintains consistent performance regardless of input language, enabling practical deployment in multilingual environments.

4) Cross-Platform Task Execution

Task execution validation demonstrates successful operation across different platforms and task types. Browser control tasks successfully open appropriate web applications and navigation interfaces. Technical documentation retrieval operates effectively through the RAG memory system and external tool integration. Information display tasks generate properly formatted HTML output for presentation on AR displays.

5) Network Architecture Validation

The network communication infrastructure successfully maintains connections across different network configurations. RTSP streaming delivers voice and video data efficiently, while RabbitMQ messaging provides reliable task coordination through proper routing and delivery acknowledgement. The system also shows resilience to network interruptions through automatic reconnection mechanisms and message persistence.

### 5. DISCUSSION

Dual-agent architecture successfully overcomes the computational challenges of merging speech recognition with intelligent processing while maintaining system modularity and scalability. Each processing stage can be optimized independently by separating concerns without compromising overall system performance.

Agent 01's dedicated speech recognition processing eliminates resource competition and provides consistent audio processing performance. The Whisper.cpp integration demonstrates effective real-time transcription capabilities with multi-language support. However, sequential processing architecture introduces cumulative latency that may impact real-time interaction requirements.

Agent 02's integration of multiple AI subsystems proves the feasibility of combining local LLM processing with external data access and memory management. The RAG memory system shows effectiveness in providing contextual information for command interpretation. The MCP tool integration successfully extends system capabilities beyond local processing limitations.

Cross-network communication capabilities address practical deployment challenges in enterprise environments. The automatic connection method detection reduces administrative overhead while maintaining appropriate security protocols. Adaptive streaming approaches balance quality and reliability across varying network conditions.

Eye tracking integration provides valuable contextual information that can enhance voice command interpretation. The binocular fusion algorithms stabilize effectively the gaze vector. Real-time data transmission via RabbitMQ maintains temporal accuracy requirements for interactive applications.

The distributed task execution framework successfully extends the AR system to AI glasses through robust messaging infrastructure. Cross-platform compatibility enables integration with existing enterprise computing environments. The modular design facilitates system expansion and component replacement without requiring architectural modifications.

The system has a few limitations. First, dependency on network connectivity for distributed task execution and computational requirements that may impact battery life during extended usage. Second, privacy concerns arise from voice data transmission and potential exposure of sensitive command information to network infrastructure.

## 6. LIMITATIONS AND FUTURE WORK

The current system implementation has several limitations that provide directions for future research. Network dependency remains a critical constraint, as distributed task execution requires stable connectivity for optimal performance. Battery consumption during extended usage sessions impacts practical deployment, particularly for mobile AR applications requiring all-day operation.

Privacy considerations arise from voice data transmission across network infrastructure, necessitating enhanced encryption and on-device processing capabilities. The sequential processing architecture, while providing modularity, introduces cumulative latency that may limit real-time interaction scenarios requiring sub-200ms response times.

Future work should focus on edge computing integration to reduce network dependency and improve privacy protection. Implementation of federated learning approaches could enable model improvement while maintaining data locality. Advanced power optimization techniques, including dynamic model switching and predictive resource management, could extend operational duration.

Multi-modal interaction combining voice, gaze, and gesture inputs represents a promising research direction for more intuitive AR interfaces. Integration with emerging 5G and edge computing infrastructure could significantly reduce latency while enabling more sophisticated AI processing capabilities.

## 7. CONCLUSION

This research demonstrates a practical approach to intelligent AI glasses systems through a multi-agent architecture and distributed processing capabilities. The system achieves functional integration of speech recognition, AI-driven processing, and task execution across diverse operational scenarios.

The dual-agent design successfully overcomes computational challenges while maintaining system modularity and scalability. Cross-network communication capabilities enable practical deployment across enterprise environments with automatic adaptation to varying connectivity conditions. In addition, multilingual processing support enables operation in diverse linguistic environments.

The distributed task execution framework extends the AR system capabilities beyond the limits of local processing by reliable messaging infrastructure. Cross-platform compatibility enables integration with existing computing environments without the need for specialized hardware deployments.

Future research directions include investigation of edge computing approaches to reduce network dependency, implementation of privacy-preserving techniques for voice data protection, and exploration of multimodal interaction combining voice, gaze, and gesture inputs. Power optimization remains a critical issue for mobile AR applications that need to operate for extended periods.

The system architecture establishes a foundation for advanced AI glasses applications that requires sophisticated voice interaction capabilities together with access to distributed computing resources. Its modular design facilitates continued development as AR hardware capabilities and AI processing techniques continue to evolve.

## REFERENCES

[1] Aouam, T., Sahnoun, S., & Hamid, M. (2018). "Voice-based Augmented Reality Interactive System for Car's Components Assembly." IEEE International Conference on Computer and Communication Engineering Technology.

[2] Fortino, G., Russo, W., Savaglio, C., Shen, W., & Zhou, M. (2017). "Agent-oriented cooperative smart objects: From IoT system design to implementation." IEEE Transactions on Systems, Man, and Cybernetics: Systems, 48(11), 1939-1956.

[3] Hosseini, M., & Georganas, N. D. (2007). "Enhanced distributed streaming system based on RTP/RTSP in resurgent ability." IEEE International Conference on Advanced Information Networking and Applications.

[4] Park, H. M., Lee, S. H., & Choi, J. S. (2008). "Wearable Augmented Reality System Using Gaze Interaction." Proceedings of the 7th IEEE/ACM International Symposium on Mixed and Augmented Reality (ISMAR '08).

[5] OpenAI. (2022). Whisper: Robust Speech Recognition via Large-Scale Weak Supervision. arXiv preprint arXiv:2212.04356. Available: https://github.com/openai/whisper

[6] FFmpeg Developers. (2023). FFmpeg: A complete, cross-platform solution to record, convert and stream audio and video. Software Documentation. Available: https://ffmpeg.org/

[7] Ollama Team. (2023). Ollama: Get up and running with large language models locally. GitHub Repository. Available: https://github.com/ollama/ollama

[8] Anthropic. (2024). Model Context Protocol: A universal protocol for connecting AI models to data sources and tools. Technical Specification. Available: https://github.com/modelcontextprotocol/

[9] ChromaDB. (2023). Chroma: The AI-native open-source embedding database. GitHub Repository. Available: https://github.com/chroma-core/chroma

[10] Reimers, N., & Gurevych, I. (2019). Sentence-BERT: Sentence embeddings using Siamese BERT-networks. Proceedings of the 2019 Conference on EMNLP, 3982-3992. Available: https://github.com/UKPLab/sentence-transformers

[11] VMware Tanzu. (2023). RabbitMQ: Open source message broker that originally implemented the Advanced Message Queuing Protocol. Software Documentation. Available: https://github.com/rabbitmq/rabbitmq-server